\documentstyle[aaspp4,tighten]{article}

\begin{document}

\title{What Can the Accretion Induced Collapse of White Dwarfs 
Really Explain?}
\author{Chris L. Fryer}
\affil{Lick Observatory, University of California at 
Santa Cruz, Santa Cruz, CA 95064}
\authoremail{cfryer@as.arizona.edu}
\author {Willy Benz}
\affil{Physikalisches Institut, Universitaet Bern, Sidlerstrasse 5, CH-3012 
Bern Switzerland}
\author{Marc Herant}
\affil{Washington University School of Medicine, Box 8107, 660 S. Euclid, 
St. Louis, MO 63110}
\author{Stirling A. Colgate}
\affil{Theoretical Division, T-6, MS B288, Los Alamos National Laboratory, 
Los Alamos, NM 87545}

\begin{abstract}

The accretion induced collapse (AIC) of a white dwarf into a neutron 
star has been invoked to explain gamma-ray bursts, Type Ia supernovae, 
and a number of problematic neutron star populations and specific binary 
systems.  The ejecta from this collapse has also been claimed as a source of 
r-process nucleosynthesis.  So far, most AIC studies have focussed on 
determining the event rates from binary evolution models and less 
attention has been directed toward understanding the collapse itself.  
However, the collapse of a white dwarf into a neutron star is followed 
by the ejection of rare neutron-rich isotopes.  The observed abundance of 
these chemical elements may set a more reliable limit on the rate at which 
AICs have taken place over the history of the galaxy.

In this paper, we present a thorough study of the collapse of a massive 
white dwarf in 1- and 2-dimensions and determine the amount 
and composition of the ejected 
material.  We discuss the importance of the input physics (equation 
of state, neutrino transport, rotation) in determining these quantities.  
These simulations affirm that AICs are too baryon rich to produce 
gamm-ray bursts and do not eject enough nickel to explain Type Ia 
supernovae (with the possible exception of a small subclass of 
extremely low-luminosity Type Ias).  Although nucleosynthesis constraints 
limit the number of neutron stars formed via AICs to $\lesssim$0.1\% of 
the total galactic neutron star population, AICs remain a viable 
scenario for forming systems of neutron stars which are difficult 
to explain with Type II core-collapse supernovae.

\end{abstract}


\section{Introduction}

White dwarfs accreting up to the Chandrasekhar limit follow one of 
two paths.  Either the densities and temperatures become sufficiently 
high to ignite explosive nuclear burning, disrupting the white 
dwarf in what is now considered the favored Type Ia explosion 
mechanism (see Woosley \& Weaver 1986 for a summary), or 
electron capture reduces central temperatures and pressures 
and instead drives a collapse of the white dwarf.  This collapse 
leads to the formation of a neutron star and is similar to the 
core collapse of massive stars, the mechanism behind Type II 
supernovae.  Just as the core collapse of massive stars ejects 
material, one might also expect material to be ejected from the 
accretion induced collapse (AIC) of white dwarfs.

Assuming some mass ejection occurs during collapse, AICs have been proposed 
as an alternate mechanism for Type Ia supernova (Colgate, Petschek, 
\& Kriese 1980) and as a source for gamma-ray bursts (Paczynski 1986, 
Goodman 1986; Goodman, Dar, \& Nussinov 1987; Paczynski 1990; Dar et 
al. 1992).  Neutron stars formed through AICs have been used to 
explain a variety of troublesome neutron star systems (see Canal, Isern, 
\& Labay 1990 for a review).  AICs have been proposed as an 
alternate channel to form neutron stars in globular clusters 
and in the galactic disk, the most common being millisecond pulsars 
(Bailyn \& Grindlay 1988; Bailyn \& Grindlay 1990; Kulkarni, 
Narayan, \& Romani 1990; Ray \& Kluzniak 1990; Ruderman 1991; 
Chen \& Ruderman 1993; Chen \& Leonard 
1993). AICs have also been invoked in several X-ray binary formation 
scenarios (Canal et al. 1990, van den Heuvel 1984) and as a formation 
mechanism for specific cases of close neutron star binaries (Ergma 1993).  

The role AICs play to produce these objects depends upon their rate.  
Super-soft X-ray sources are possible candidates 
of white dwarfs accreting up to the Chandrasekhar limit (Li \& 
van den Heuvel 1997).  However, whether or not the white dwarf 
will form a Type Ia supernova in a thermonuclear explosion 
or collapse into a neutron star in an AIC depends sensitively 
upon the initial white dwarf mass, white dwarf composition, 
and the accretion rate onto the white dwarf (Nomoto 1982, 1984; 
Nomoto \& Kondo 1991).  Although current observations verify that 
likely progenitors for AICs do exist, they do not place strong 
quantitative constraints on the event rate of these collapses.  
Similarly, uncertainties in binary evolution and white dwarf 
formation make it difficult to predict any definitive AIC event 
rate from population synthesis calculations (Yungelson \& Livio 1998).  

By simulating the collapse of white dwarfs, and their subsequent 
explosions, we can constrain the viability of AICs as 
gamma-ray bursts and Type Ia supernovae mechanisms.  In addition, 
we can also use the nucleosynthetic yield from the 
ejecta of AICs to place limits on the event rate.  The ejecta of AICs 
is neutron rich and leads to the production of many anomalous 
neutron-rich isotopes (e.g. $^{62}$Ni, $^{66}$Zn, $^{68}$Zn, $^{87}$Rb, 
and $^{88}$Sr) which pollute the interstellar medium. By comparing the 
observed abundance of these elements with the amount ejected per AIC event, 
we can place constraints on the allowable rate of AICs in the galaxy 
(Woosley \& Baron 1987, hereafter WB87).   

Previous work on AICs has identified three possible mass-ejection 
mechanisms:  the prompt mechanism driven by the bounce 
of the collapsing white dwarf as the core reaches nuclear densities, 
the ``delayed-neutrino'' mechanism which occurs shortly after the 
bounce-shock stalls (20-200 ms) and is driven by neutrino absorption, 
and the neutrino wind mechanism which is a relatively stable 
mass-loss occurring over the relatively long cooling timescales (1-2\,s) 
of the proto-neutron star.  Baron et al. (1987), 
Mayle \& Wilson (1988) and  Woosley \& Baron (1992) 
all found that the bounce shock stalls due to the energy losses 
from neutrino emission and dissociation and the prompt mechanism 
fails to drive an explosion.  Mayle \& Wilson's simulations of the 
collapse of massive star ($8-10 M_{\odot}$) OMgNe cores, which have similar 
structures to most AIC progenitor models, led to 
explosions on short timescales ($\sim 200$\,ms) via the delayed-neutrino 
mechanism with $0.042 M_{\odot}$ ejected.  Simulations by 
Hillebrandt, Nomoto, \& Wolff (1984) of the collapse of OMgNe 
white dwarfs ejected as much as $\sim 0.1-0.2 M_\odot$ with 
explosions which developed even sooner (20-30\,ms).
These explosions occur so quickly because the collapsing white 
dwarf does not have a huge infalling mantle that provides 
a ram pressure containing, at least temporarily, the explosion.  
WB92, however, found that no such 
explosion resulted from the collapse of a CO white dwarf.  
The only mass-loss ($\sim 0.01 M_{\odot}$) occurred at 
late times through the proto-neutron star's neutrino-driven wind.  

In both the delayed-neutrino explosion mechanism and the 
neutrino-driven wind, the material ejected is likely to be 
neutron rich.  Shortly after the inception of the neutrino-driven 
supernova mechanism, it was realized (Arnett \& Truran 1970) that 
the densities and temperatures near the neutrinosphere 
are sufficiently high to force this material to deleptonize 
via the emission of electron neutrinos.  
This material is ejected with extremely low electron fractions 
($0.35<Y_e<0.45$).  In Type II supernovae simulations, the 
assumption is that either a longer delay in the supernova 
explosion (due to the ram-pressure of the infalling material) 
causes most of this material to remain part of the neutron 
star, or that the neutron-rich material falls back on the neutron star.   
The fallback is driven by the reverse shock that is created as the 
supernova shock wave traverses the envelope of the massive star. 
As we shall show in this paper, neither of these arguments can possibly 
hold for AICs and we can not easily explain away the low $Y_e$ 
ejecta from the delayed neutrino mechanism.  

Neutrino {\it emission} is not the only way to lower the 
electron fraction of matter.  The electron fraction neutrino-wind 
driven ejecta is set by the relative absorption of electron neutrinos and 
anti-neutrinos (Qian et al. 1993).  Qian et al. (1993) have shown 
that since the neutrinosphere of the electron anti-neutrinos 
is deeper within the proto-neutron star crust, the anti-neutrinos 
are more energetic than the electron neutrinos.  Since the 
neutrino cross-section is proportional to the square of the neutrino 
energy, given the same flux of neutrino/anti-neutrinos, 
the wind driven material is likely to absorb more anti-neutrinos 
and it becomes neutron rich. 

In this paper, we give results from a series of AIC simulations using the CO 
white dwarf progenitor from the WB92 model 
to determine the amount and composition of the ejecta.  To measure 
the reliability of our results, we vary a number of parameters
such as the details of the neutrino physics, the equation of state, 
and the initial rotation of the white dwarf.  In particular, we 
are able to understand the difference between the WB92 results 
and the other groups.  We discuss the models 
in detail in \S 2.  A summary of these results and their 
implications are given in \S 3.  

\section{Models and Results}

Table 1 summarizes the entire set (60 in total) of 
simulations we have performed. The different simulations 
were run to test the sensitivity of the results to changes in the 
neutrino physics (both source and transport columns 3 and 4 of Table 1), 
the inclusion of relativistic effects, (runs 6, 11, and 17), the
choice of the equation of state (EOS) (column 2), multi-dimensional 
effects (run 2) and initial rotation rate of the
white dwarf (run 3). 

Except for changes in the equation of state for dense matter, most 
of these parameter variations lead to relatively small changes 
in the results (factors of 2 in the mass ejected).  Changes in 
the equation of state explain the differences between the 
previous simulations (Hillebrandt, Nomoto \& Wolff 1984 and   
Mayle \& Wilson 1988 versus Woosley \& Baron 1992).  We include 
calculations using both the equation of state of Swesty \& Lattimer
(1992) and that of Baron, Cooperstein, \& Kahana (1985). The large differences 
can be appreciated by comparing the mass-point trajectories over 
the course of the simulation (Figs 1,2).  In this section we 
discuss the specific variations in our simulations and their 
effects on the results.  

To calculate the upper limit of the event rate of AICs, we must estimate 
the nucleosynthetic yield of the neutron rich ejecta.  
Hartmann, Woosley, and El Eid (1985) estimate that there must be less than 
$10^{-5} M_{\odot}$ of $Y_e < 0.4$ material ejected per supernova to avoid 
anomalous abundances of particular isotopes (e.g. $^{62}$Ni, $^{66}$Zn, 
$^{68}$Zn, $^{87}$Rb, and $^{88}$Sr).  Using the value of $0.02$ 
$M_{\odot}$ of material with less than $Y_e < 0.4$ ejected per AIC events 
(see Table 1) and assuming a supernova rate of two per century for the 
Galaxy we find that the upper limit for rate of AICs must be $(2/100 {\rm 
y}^{-1})(10^{-5} M_{\odot})/(0.02 M_{\odot}) = 10^{-5}$ y$^{-1}$. Note 
that we have assumed AICs to be the only source for material with such a 
low $Y_e$. Should there be another source of these neutron rich isotopes, 
the allowed AIC rate will be correspondingly smaller.  
A similar constraint can be calculated by using the material ejected 
with $0.45 < Y_e < 0.40$ rather 
than with $Y_e < 0.4$. Following the method of WB92 for $0.02 M_{\odot}$ 
of ejected material, the upper limit for the rate of AICs becomes 
$(1 M_{\odot}/170,000)(1/0.02 M_{\odot}^{-1})(1/0.13)
(2/100{\rm y}^{-1}) = 4.5 \times 10^{-5}$ y$^{-1}$.  
In table 1, we list the upper limit of the event rate for each simulation.  In 
parentheses, we list the same upper limit if the electron fraction 
of the ejecta is 30\% higher than the predictions in our simulations.  
We note that much of the ejecta has a very low electron fraction 
($Y_e<0.3$) and large errors in the estimated electron fraction 
would be required to change the upper limit of the AIC event 
rate by more than an order of magnitude.

\subsection{Numerical Methods}

The internal structure of the initial white dwarf is taken from the 
progenitor used by WB92.  This model is then mapped into 
our one- and/or two-dimensional codes and run for $0.2$s. The 
one-dimensional simulations were performed using the 
code developed and tested in previous work (Benz 1991;  
Herant et al. 1994; Fryer, Benz \& Herant 1996) with 
$\sim 110$ zones where the highest resolution was constructed 
near the mass cut. This code does not include 
any form of convection modeling (mixing length or other). 
The two-dimensional simulations were performed using the Smooth Particle 
Hydrodynamics (SPH) code discussed in Herant et al. (1994) with 
typically $\sim 8000$ particles.  We model a 180$^\circ$ wedge assuming 
cylindrical symmetry about the angular momentum axis and also 
run these models for $0.2$s.  As described in Herant et al. (1994), 
both codes use the same implementations of neutrino physics and 
transport, equations of state, etc.  Thus, we can use the 
two dimensional simulations to compare the effects of convection 
and rotation.

For some of the one-dimensional simulations, we include general 
relativistic effects using the formalism developed by van Riper (1997).  
For our late-time simulations, we have added a cell-splitting 
routine which allows us to follow the evolution of the 
explosion long after the collapse (we have run select simulations 
to 0.5\,s).  We only use the cell-splitting routine at late 
times, after the explosion has occurred.  When the cell size 
becomes a sizable fraction (0.3) of its radius, we divide 
the cell in half and reduce the energy (and hence pressures) 
of the inner cell by 5\% to allow the forces at the boundaries 
to remain roughly equal.

\subsection{Effects of Neutrino Physics}

To illustrate the importance of neutrinos on the composition of the 
ejecta, we compare the results of a simulation which includes the effects 
of neutrino physics (Fig. 1 - run 1) and a simulation with no neutrino 
emission or absorption (Fig. 3 - run 24).  
By comparing the mass-point trajectories between 
these two simulations, we note that without the cooling effects of neutrino 
emission, the bounce shock does not stall and an explosion develops 
(a ``prompt'' explosion).  For the simulation which includes the 
effects of neutrino physics, 
neutrino emission from the shocked material (along with dissociation) 
stalls the shock.  The neutrino emission from the shocked material 
and the new material falling onto the shocked region serves to 
deleptonize the material.  Using the Swesty-Lattimer (1992) equation 
of state, we find that neutrino heating is able to revive the explosion, 
ejecting $\sim 0.1 M_{\odot}$ of material.  The neutrino emission 
and later absorption sets the electron fraction of the material.  
However, very quickly, the ejecta is thrown sufficiently far 
where adiabatic cooling causes recombination which lowers the 
free proton and free neutron fraction.  The neutrino opacity 
of this material drops, and the electron fraction is effectively 
``frozen-out''.  In all 
cases where the delayed-neutrino mechanism is the dominant mass 
ejector, we follow our simulations until this occurs.  

Including neutrino physics in the simulations involves two 
difficulties: the determination and numerical representation of
the processes that emit or absorb neutrinos and the subsequent
transport of these neutrinos through matter. 
Our neutrino processes are described in Herant et al. (1994) and 
include many of the possible emission and absorption rates for the 
standard three neutrino species (electron neutrino, electron antineutrino, 
and the entire set of $\mu$ and $\tau$ neutrinos and antineutrinos).  

In many Type II supernova simulations, it is often assumed that 
the neutrinos are emitted from ultra-relativistic electrons ($T \gg 
\frac{1}{2} MeV$). This is not always necessarily a good approximation.
An analytical approximation of the electron/positron neutrino 
emission rates and luminosity has been derived by Takahashi, El Eid 
\&Hillebrandt (1977):
\begin{eqnarray}
\lambda_{\stackrel{e^{-}p}{e^{+}n}} = C_2
\beta^{-5} \left[ \rule{0mm}{6mm}
F_4(\pm \eta) \pm (2-2 \Delta) F_3(\pm \eta) \beta +
\left( \frac{1 - 8 \Delta + 2 \Delta^{2}}{2} \right)
F_2(\pm \eta) \beta^2 \right. \nonumber \\ \left. \pm (2 \Delta^2 - \Delta)
F_1(\pm \eta) \beta^3 + \left( \frac{4 \Delta^{2}-1}
{8} \right) F_0(\pm \eta) \beta^4 \right]
\end{eqnarray}
and
\begin{eqnarray}
L_{\stackrel{\nu}{\bar{\nu}}}=C_3 \beta^{-6} \left[ \rule{0mm}{6mm}
F_5(\pm \eta) \pm (2-3 \Delta) F_4(\pm \eta) \beta +
\left( \frac{1 - 12 \Delta + 6 \Delta^{2}}{2} \right)
F_3(\pm \eta) \beta^2 \right. \nonumber \\ \left.
\pm \left( \frac{-3 \Delta \pm
12 \Delta^2 - 2 \Delta^{3}}{2} \right) F_2(\pm \eta) \beta^3 +
\left( \frac{-1 + 12 \Delta^2 - 16 \Delta^{3}}{8} \right)
F_1(\pm \eta) \beta^4 \right. \nonumber \\ \left.
\pm \left( \frac{2 + 3 \Delta - 4
\Delta^{3}}{8} \right) F_0(\pm \eta) \beta^5 \right]
\end{eqnarray}
where $\lambda_{\stackrel{e^{-}p}{e^{+}n}}$ are the transition
rates for electron and positron capture, $L_{\stackrel
{\nu}{\bar{\nu}}}$ are the electron neutrino and anti-electron
neutrino luminosities, $C_2=6.15 \times 10^{-4}$ s$^{-1}$ per nucleon,
$C_3=5.04 \times 10^{-10}$ erg s$^{-1}$ per nucleon, $\Delta = 1.531$ MeV
is the neutron-hydrogen mass difference and $\beta = \frac{m_e c^2}
{k_B T}$.  $F_n$ are fermi integrals of order n and $\eta$ is the
degeneracy parameter.  

Adopting the ultra-relativistic limit, WB92 have simplified these
equations, taking only the first term in each equation. To estimate
the importance of this assumption, we have used both the limited and 
the full equations. In Table 1, the runs using the full equations are
identified by the letters TEH while those using the ultra-relativistic
assumption are marked by WB. As can be seen from a careful comparison 
of these simulations, adopting the ultra-relativistic limit 
changes the amount of neutron rich ejecta (and the corresponding 
AIC event rate) by less than a factor of 2.

Despite the low cross-sections for neutrino interactions, the high 
densities involved in core-collapse scenarios place the neutrinos 
within the depths of the collapsing star in the diffusion regime.  
Thus, neutrino transport must include both the diffusion and 
free-streaming limits of the transport equations. The ``standard'' 
approximation to couple these two extremes calls upon the use of a 
flux-limiter (see for example, Janka 1991). We have incorporated 
several different flux-limiters (Bowers \& Wilson 1982; Levermore 
\& Pomraning (1981);  Herant et al. 1994) and the properties of their 
ejecta can be compared in column 2 of Table 1.  The Bowers-Wilson 
and Levermore-Pomraning flux limiters seem to bound the more accurate 
Monte-Carlo calculations by Janka (1991) and can be used to gauge the 
effect of the flux-limiter on the amount and composition of the ejecta. 
By comparing these two flux limiters in otherwise identical 
simulations (e.g. run 8 and 12), we see that the two approximations 
in the neutrino diffusion lead to only 10\% differences in the mass 
ejected.  The upper limit on the AIC rate with these two flux 
limiters varies by factors of 2.

\subsection{Impact of the Equation of State}

Uncertainties surrounding the equation of state for dense matter 
lead to the largest differences in the ejecta from AICs.  
We use two such equations of state:  the one described in Herant 
et al. (1994) which couples the nuclear equation of state by Lattimer 
\& Swesty (1991) to a low density equation of state 
(Blinnikov, Dunina-Barkovskaya \& Nadyozhin 1996) and a nuclear statistical 
equilibrium (NSE) scheme (Hix et al. 1994) (hereafter called SL EOS); and the 
equation of state developed by Baron, Cooperstein, \& Kahana (1985 hereafter 
called BCK EOS) which covers both low and high density regimes.  
The BCK EOS is the equation of state used by WB92. For our progenitor, 
the effects of nuclear 
burning were minimal. We verified this by running a simulation (run 7) 
in which we calculate the energy from  nuclear burning with a 14 
element nuclear network (Benz, Hills, \& Thielemann 1989) rather 
than assuming nuclear statistical equlibrium.  As with the neutrino 
physics, nuclear burning varies the critical AIC rate by factors of 2 only.  

The main results of our simulations using one of the two 
equations of state are listed in Table 1 (see column 1 - e.g. compare 
runs 8 and 20). By using the BCK equation of state, we recover the results 
of WB92.  The equation of state dramatically affects the amount 
and compositon of the ejecta in the simulations. Swesty, Lattimer, 
\& Myra (1994), in a previous comparison between the two equations of 
state, report similar findings (the softer BCK equation of state leads 
to denser, and hotter, cores after bounce). These differences have been 
discussed by Swesty, Lattimer, \& Myra who argue that, given the 
standard equation of state parameters for the BCK EOS, the SL EOS is 
physically more accurate. We have run the SL EOS using two values for 
the incompressibility of bulk nuclear matter ($K_s = 180,375$ MeV) and 
have run a grid of the faster BCK EOS varying the BCK gamma ($1.5 < 
\gamma < 3.5$), the bulk surface coefficient ($25 < W_s < 38$), the 
symmetric bulk compressibility parameter ($180 \,{\rm MeV} < K^{sym}_0 < 
375 \, {\rm Mev}$), 
and an asymmetry parameter ($1.5 < xkz < 3.5$).  Table 2 gives the 
results for this grid of simulations in which we use the Levermore-Pomraning 
flux limiter.  Despite the wide range in the physical parameters, 
the results from the BCK EOS never agree with those from the SL EOS.  
Clearly, the differences between the two equations of state goes 
beyond compressibility or asymmetry parameters.  

The ratio of the SL EOS pressure to the BCK EOS pressure along an 
$S=2$ k$_B/$nucleon adiabat is plotted in Figure 4. Note that for 
densities less than $10^{14}$ g cm$^{-3}$, the pressure of the 
BCK EOS is 10-20\% greater than that of the SL EOS.  Relatively 
small differences such as these mark the difference between 
a success or failure of the delayed-neutrino explosion mechanism, 
which then leads to greater than order-of-magnitude differences 
in the ejecta!

We have also studied the effects of general relativity on the 
simulations using the SL equation of state.  From Table 1, 
we can compare the results of simulations with or without general 
relativity (runs 1,6). The primary effect of general relativity is 
to cause the 
material to fall deeper into the potential well resulting in increased 
neutrino emission/absorption.  Just as the equation of state strongly 
affects the amount and composition of the ejecta, the addition of 
general relativity leads to variations of over an order of magnitude 
in the upper limit of AIC event rate.

\subsection{Effects of Convection and Rotation}

Two-dimensional simulations of AICs can be used to test both the 
effects of convection and rotation.  Since large entropy gradients 
do not develop in the one-dimensional simulations, we do not 
expect convection to cause large differences in the ejecta from AICs.  
We first use our two-dimensional simulations to verify that convection 
does not play a major role in the collapse and ejecta of AICs.  
Although convection is indeed present in the simulations (Fig. 5), 
the explosion occurs so rapidly ($<100$\,ms) that it has no 
time to alter the explosion results.  The small differences between 
the one- and two-dimensional models (compare run 1 and run 2 in Table 1) 
are probably entirely due to the resolution differences between the 
two simulations.  

The progenitors to AICs accrete not only mass, but angular momentum, 
as the white dwarf approaches the Chandrasekhar limit.  Typical rotation 
periods for cataclysmic variables range from $200-1200$ s (Liebert 1980) 
although periods of $\sim 30$ s exist (King \& Lasota 1991).  A lower limit 
on the rotation period is set by the break-up spin period ($\sim 0.5$ s 
for solar-mass white dwarfs).   In our models, we assume solid-body 
rotation and conserve each individual particle's angular momentum 
for the duration of the simulation.  We take an extreme case of a 
white dwarf rotating with a 20\,s period prior to collapse (Fig. 5).  

For this short rotation period, the ratio of surface rotational velocity 
over the keplerian velocity of the outer material exceeds 0.1 as the 
white dwarf collapses.  This will certainly alter the flow of the 
outer material.  However, since white dwarfs are probably solid-body 
rotators, the bulk of the inner material is unaffected by rotation.  
For example, the ratio of rotational over keplerian velocity drops to 
$0.01$ at the radius which encloses $1.2 M_{\odot}$ (still 87 \% of 
the white dwarf mass). Thus, while the collapse of the outer envelope 
of a white dwarf is affected by rotation, the core collapse itself 
is not (compare T=50,\,70\,ms in Fig. 5). Since most of the material 
exterior to $1.2 M_\odot$ is ejected, the net effect of rotation 
is negligible.  Comparing runs 2 and 3 in Table 1, we see that 
the amount and composition of the ejecta is not altered significantly 
by the effects of rotation.  In addition, there is no preferential 
ejection of material along either axis.

We do not follow the proto-neutron star as it cools and contracts 
and the spin period at the end of our simulation (where the hot 
proto-neutron star's radius is $30-50$\,km) is still $\sim 1$\,s.  
Although much of the white dwarf's angular momentum is ejected with 
the outer $0.2 M_\odot$, as the proto-neutron star continues to 
contract down to 10\,km, unless further material is ejected, 
its spin period will decrease to 10\,ms.  This system is likely 
to continue to accrete from the same companion that caused it 
to collapse in the first place and this may speed up the 
neutron star's spin even further, producing millisecond 
spin periods.

\subsection{Neutrino Wind}

The results in Table 1 do not include any mass loss from neutrino-driven 
winds.  The amount of this ejecta is small when compared to that of our 
simulations using the SL equation of state (WB92 predict 
$\sim 0.005M_{\odot}$\,s$^{-1}$ for the first two seconds).  
However, for the BCK EOS simulations, neutrino-driven 
winds dominate the mass loss.  Using a cell-adding routine, we follow the 
fate of an AIC 
model beyond the delayed neutrino-induced explosion to obtain the wind 
driven mass loss (run 1). Because we are also 
modeling the core, we are limited to very small timesteps and are
able to follow this phase for only $0.5$s after bounce. We find that
during this period of time, an additional $\sim 0.002 M_{\odot}$ 
peels off the neutron star. Thus, in this limited way, we confirm the
results obtained by WB92 that further neutrino driven mass loss is to be 
expected. Independent of the equation of state or other input physics, it 
is likely that a neutrino-wind phase exists and that $\sim 0.01 M_{\odot}$ 
is ejected during this phase.  However, the electron fraction of 
such a wind is very sensitive to the flux and energy of the 
electron neutrino/anti-neutrinos.  We do not list the upper 
limits for the AIC event rate placed by ``wind'' ejecta, but 
prefer to rely upon the ejecta from the delayed-neutrino 
mechanism which we are better able to test with our simulations.

\section{So What Are AIC Good For?}

Table 1 summarizes the results of our suite of accretion-induced 
collapse simulations varying the input physics within 
the range of current uncertainties.  We find that the differences 
in past work were primarily due to differences in the 
equations of state.  Even with these uncertainties, the 
study of explosions resulting from the collapse of white 
dwarfs, can help determine the viability of AICs as models 
for Type Ia supernovae and gamma-ray bursts.  In addition, by 
calculating the nucleosynthetic yields of the neutron rich ejecta 
from AICs, we can limit the event rate (following the procedure described 
in \S 2).  The upper limit of the AIC rate is given in Table 1 and, 
if we use the possibly more reliable Lattimer \& Swesty (1991) equation 
of state, is roughly $10^{-7}-10^{-5}$\,yr$^{-1}$.  To change 
these timescales appreciably, there must be large changes in the 
electron fraction of the ejecta which, although unlikely, 
can not be excluded at this time.  

\subsection{Neutron Star Populations}
Our upper limit on the event rate somewhat constrains 
the role AICs play in producing neutron stars.  The Type II 
supernova rate ($\sim 10^{-2}$\,yr$^{-1}$:  Cappellaro et al. 1998) 
is 3-4 orders of magnitude higher than our upper limit of the AIC rate, 
so neutron stars formed from AICs will not make up a large fraction 
of the total population of galactic neutron stars.  However, neutron 
stars formed via accretion induced collapse may not receive the 
same large kicks observed in neutron stars formed in Type II 
supernovae, and hence may be prime candidates for globular 
cluster neutron star populations.  Bailyn \& Grindlay (1990) 
estimate that the AIC rate must be $\sim 10^{-5}-10^{-5}$\,yr$^{-1}$ 
to explain the globular cluster neutron star population, barely 
consistent with our upper limit on the AIC event rate.  In 
addition, the event rate does not preclude AICs explaining 
any peculiar neutron star systems.

\subsection{Nucleosynthesis}
Because of their neutron rich ejecta, AICs are prime sites for 
r-process nucleosynthesis (Wheeler, Cowan, \& Hillebrandt 1998).  
However, although we can constrain the AIC event rate from 
the nucleosynthetic yield with our current models, predicting 
yields of specific isotopes remains beyond our grasp.  To predict 
precisely the nucleosynthetic yields from AICs, the physical 
processes that cause the largest variations in the results
(in particular, the equation of state and the effects of 
general relativity, and possibly, the characteristics 
of the progenitor) must be precisely known.

\subsection{Gamma-Ray Bursts}
To achieve the high Lorentz factors required to drive a gamma-ray 
burst, a viable mechanism must have both large explosion energies 
($\sim 10^{51}$\,ergs for isotropic explosions) and eject 
very little mass ($\lesssim 10^{-5} M_\odot$).  For our 
most-likely models, AICs eject 4 orders of magnitude too much 
mass with an order of magnitude too little energy.  In agreement 
with Woosley \& Baron (1992), this clearly rules out those gamma-ray 
burst models relying upon neutrino/anti-neutrino annihilation
(Paczynski 1986; Goodman 1986; Goodman, Dar, \& Nussinov 1987; 
Paczynski 1990; Dar et al. 1992).  It also rules out all 
magnetically beamed models of AICs (Shaviv \& Dar 1995; 
Yi \& Blackman 1997,1998; Dai \& Lu 1998).  In beamed models, 
the actual ejecta that effects the gamma-ray burst is limited 
to the ejecta swept up in the beam.  In order for 
these beamed models to avoid ejecting too much mass, 
the beaming fraction must be smaller than 0.01\% of the sky 
(hence sweeping up only 0.01\% of the mass).  
However, with such a small beaming fraction, the AIC event rate 
must be greater than $10^{-3}$\,yr$^{-1}$ for this mechanism 
to make up the majority of the observed gamma-ray bursts.  This 
is an order of magnitude above our upper limit for the AIC 
event rate and it appears that AICs are simply unable 
to meet the observed requirements of gamma-ray bursts.

\subsection{Type Ia Supernovae}
On the other hand, AICs do not eject {\it enough} nickel to 
match most Type Ia supernova light curves.  The amount 
of nickel ejecta is tantalizingly close to the properties of some 
peculiar Type Ia supernovae with very low nickel masses 
(e.g. SN 1991bg: Filippenko et al. 1992).  AICs may be able 
to explain some peculiar Ias, a possibility that could be 
confirmed by either neutrino detections or the discovery
of a neutron star formed in these Ias.

\acknowledgements
This paper has benefitted from the contributions of many people.  
We are grateful to R. Hix for making available his nuclear 
statistical equilibrium code, to D. Swesty for his nuclear 
equation of state, to D. Nadyozhin for his equation of state, 
to C. Wingate for his graphics software, and to S. Woosley 
for the progenitor star and many useful discussions.  We would 
especially like to thank Edward Baron for providing access to 
his equation of state, insightful advice, and encouragement.  
The work of C.F. and W.B. was partially supported 
by NSF grant AST 9206738 and from the Swiss 
National Science Foundation.  The work of M. H. was supported 
by a director funded post-doctoral fellowship at Los 
Alamos National Laboratory.


\begin{deluxetable}{llcccccc}
\tablewidth{43pc}
\tablecaption{AIC Simulations}
\tablehead{
\colhead{Model} &
\colhead{EOS\tablenotemark{a}}      
& \colhead{Flux\tablenotemark{b}} &
\colhead{$\nu$\tablenotemark{c}} &
\colhead{} & \colhead{M$_{\rm ej} (M_{\odot})$} &
\colhead{} & \colhead{Rate\tablenotemark{d}} \\
\colhead{Num.} &
\colhead{} & \colhead{Limiter} &
\colhead{Source Term}      & \colhead{Y$_{\rm e} > 0.45$} 
 & \colhead{$0.40 < $Y$_{\rm e} < 0.45$} &
\colhead{Y$_{\rm e} < 0.40$} & \colhead{(AIC/Myr)}}

\startdata
1 & SL & H{\it etal} & TEH & 0.09 & 0.04 & 0.07 & 2.9(12.9) \nl
2 & SL-2D\tablenotemark{e} & H{\it etal} & TEH  & 0.07 & 0.01 & 0.09 & 2.2(10.1) \nl
3 & SL-2D (rot) & H{\it etal} & TEH & 0.07 & 0.01 & 0.11 & 1.8(8.2) \nl
4 & SL & H{\it etal} & WB & 0.10 & 0.02 & 0.08 & 2.5(11.3) \nl
5 & SL-375\tablenotemark{f} & H{\it etal} & TEH & 0.08 & 0.03 & 0.09 & 2.2(10.1) \nl
6 & SL-GR\tablenotemark{g} & H{\it etal} & TEH & 0.02 & 0.04 & 0.22 & 0.91(4.1) \nl
7 & SL-Adv. Burn\tablenotemark{h} & H{\it etal} & TEH & 0.07 & 0.05 & 0.05 & 4.0(18.1) \nl
8 & {\bf SL}\tablenotemark{i} & {\bf BW} & {\bf TEH} & {\bf 0.05} 
& {\bf 0.06} & {\bf 0.02} & {\bf 10.0(45.2)} \nl
9 & SL & BW & WB & 0.07 & 0.03 & 0.03 & 6.7(30.2) \nl
10 & SL-Adv. Burn & BW & TEH  & 0.04 & 0.06 & 0.005 & 15.1(181.0) \nl
11 & SL-GR & BW & TEH & 0.01 & 0.01 & 0.26 & 0.77(3.5) \nl
12 &  {\bf SL} & {\bf LP} & {\bf TEH} & {\bf 0.05} & {\bf 0.06} 
& {\bf 0.01} & {\bf 15.1(90.5)} \nl
13 & SL & LP & WB & 0.07 & 0.05 & 0.02 & 10.0(45.2) \nl
14 & SL-Adv. Burn & LP & TEH & 0.06 & 0.03 & 0.02 & 10.0(45.2) \nl
15 & SL-No Burn & LP & TEH & 0.1 & 0.07 & 0.03 & 6.7(30.2) \nl
16 & SL-No Burn & LP & WB & 0.08 & 0.05 & 0.03 & 6.7(30.2) \nl
17 & SL-GR & LP & TEH & 0.01 & 0.01 & 0.25 & 0.80(3.6) \nl
18 & BCK & H{\it etal} & TEH & 0.04 & 0.04 & 0.0 & NA \nl
19 & BCK & H{\it etal} & WB & 0.03 & 0.03 & 0.0 & NA \nl
20 & BCK & BW & TEH & 0.0 & 0.0 & 0.0 & NA \nl
21 & BCK & BW & WB & 0.0 & 0.0 & 0.0 & NA \nl
22 & BCK & BW & TEH & 0.0 & 0.0 & 0.0 & NA \nl
23 & BCK & BW & WB & 0.0 & 0.0 & 0.0 & NA \nl
24 & SL & none & none & 0.20 & 0.0 & 0.0 & $\infty$ \nl

\tablenotetext{a}{We use either the Swesty-Lattimer (SL) or
Baron-Cooperstein-Kahana (BCK) equations of state}
\tablenotetext{b}{The following flux limiters are employed:  
Herant et al. (H{\it etal}); Bowers-Wilson (BW); 
Levermore-Pomraning (LP).}  
\tablenotetext{c}{We use the detailed electron/positron emission 
rates of Takahashi, El Eid \&Hillebrandt (TEH-1977) or the simplified 
rates used by WB.}
\tablenotetext{d}{These rates are the maximum allowed assuming the entire 
production of these neutron rich materials come from AICs.  The 
number in parantheses denotes the limit on the rate if the true $Y_e$ 
of the ejecta is 10\% higher than our simulations predict.}
\tablenotetext{e}{The 2-D models are SPH simulations with $\sim 8000$ 
particles.  The rotational simulation assumes a white dwarf spinning 
with a period of 2\,s.}
\tablenotetext{f}{$K_s = 375$MeV.  In all other SL EOS simulations, $K_s
= 180$MeV.}
\tablenotetext{g}{This run includes general relativistic effects.}
\tablenotetext{h}{Some runs include an additional burning network to 
be used prior to the onset of nuclear statistical equilibrium.  In 
other runs, we have turned off even the NSE network to test 
the range of effects from nuclear burning.}
\tablenotetext{i}{The most probable outcomes are bold-faced.}
\enddata
\end{deluxetable}

\begin{deluxetable}{lcccccc}
\tablewidth{0pc}
\tablecaption{BCK EOS models\tablenotemark{a}}
\tablehead{
\colhead{$BCK \gamma$}      
& \colhead{$W_s$} &
\colhead{$K_0^{sym}$} & \colhead{$xkz$} &
\colhead{Result}}

\startdata
1.5 & 31.5 & 180 & 2.0 & -\tablenotemark{b} \nl
1.5 & 31.5 & 375 & 2.0 & - \nl
2.5 & 31.5 & 180 & 2.0 & - \nl
2.5 & 31.5 & 375 & 2.0 & - \nl
3.5 & 31.5 & 180 & 2.0 & - \nl
3.5 & 31.5 & 375 & 2.0 & - \nl
1.5 & 25.0 & 180 & 2.0 & - \nl
1.5 & 25.0 & 375 & 2.0 & - \nl
3.5 & 25.0 & 180 & 2.0 & - \nl
3.5 & 25.0 & 375 & 2.0 & - \nl
1.5 & 38.0 & 180 & 2.0 & - \nl
1.5 & 38.0 & 375 & 2.0 & - \nl
3.5 & 38.0 & 180 & 2.0 & - \nl
3.5 & 38.0 & 375 & 2.0 & - \nl
1.5 & 25.0 & 180 & 1.5 & - \nl
1.5 & 25.0 & 375 & 1.5 & - \nl
3.5 & 25.0 & 180 & 1.5 & - \nl
3.5 & 25.0 & 375 & 1.5 & - \nl
1.5 & 38.0 & 180 & 1.5 & - \nl
1.5 & 38.0 & 375 & 1.5 & - \nl
3.5 & 38.0 & 180 & 1.5 & - \nl
3.5 & 38.0 & 375 & 1.5 & - \nl
1.5 & 25.0 & 180 & 2.5 & $0.05 M_{\odot}$ ejected \nl
1.5 & 25.0 & 375 & 2.5 & - \nl
3.5 & 25.0 & 180 & 2.5 & $0.05 M_{\odot}$ ejected \nl
3.5 & 25.0 & 375 & 2.5 & - \nl
1.5 & 38.0 & 180 & 2.5 & - \nl
1.5 & 38.0 & 375 & 2.5 & - \nl
3.5 & 38.0 & 180 & 2.5 & - \nl
3.5 & 38.0 & 375 & 2.5 & - \nl
2.5 & 25.0 & 180 & 2.0 & - \nl
2.5 & 25.0 & 375 & 2.0 & - \nl
3.5 & 25.0 & 180 & 2.0 & - \nl
3.5 & 25.0 & 375 & 2.0 & - \nl
2.5 & 25.0 & 180 & 2.5 & $0.05 M_{\odot}$ ejected \nl
2.5 & 25.0 & 375 & 2.5 & - \nl
2.5 & 31.5 & 180 & 2.5 & - \nl
2.5 & 31.5 & 375 & 2.5 & - \nl

\tablenotetext{a}{\footnotesize In all these models we use the Levermore-Pomraning 
flux limiter and the full neutrino source terms.}
\tablenotetext{b}{\footnotesize Except where specifically noted, the end result 
for these simulations was no explosion.  Of course, this does not 
exclude mass ejection from neutrino winds.}
\enddata
\end{deluxetable}


\clearpage

\begin{figure}
\plotfiddle{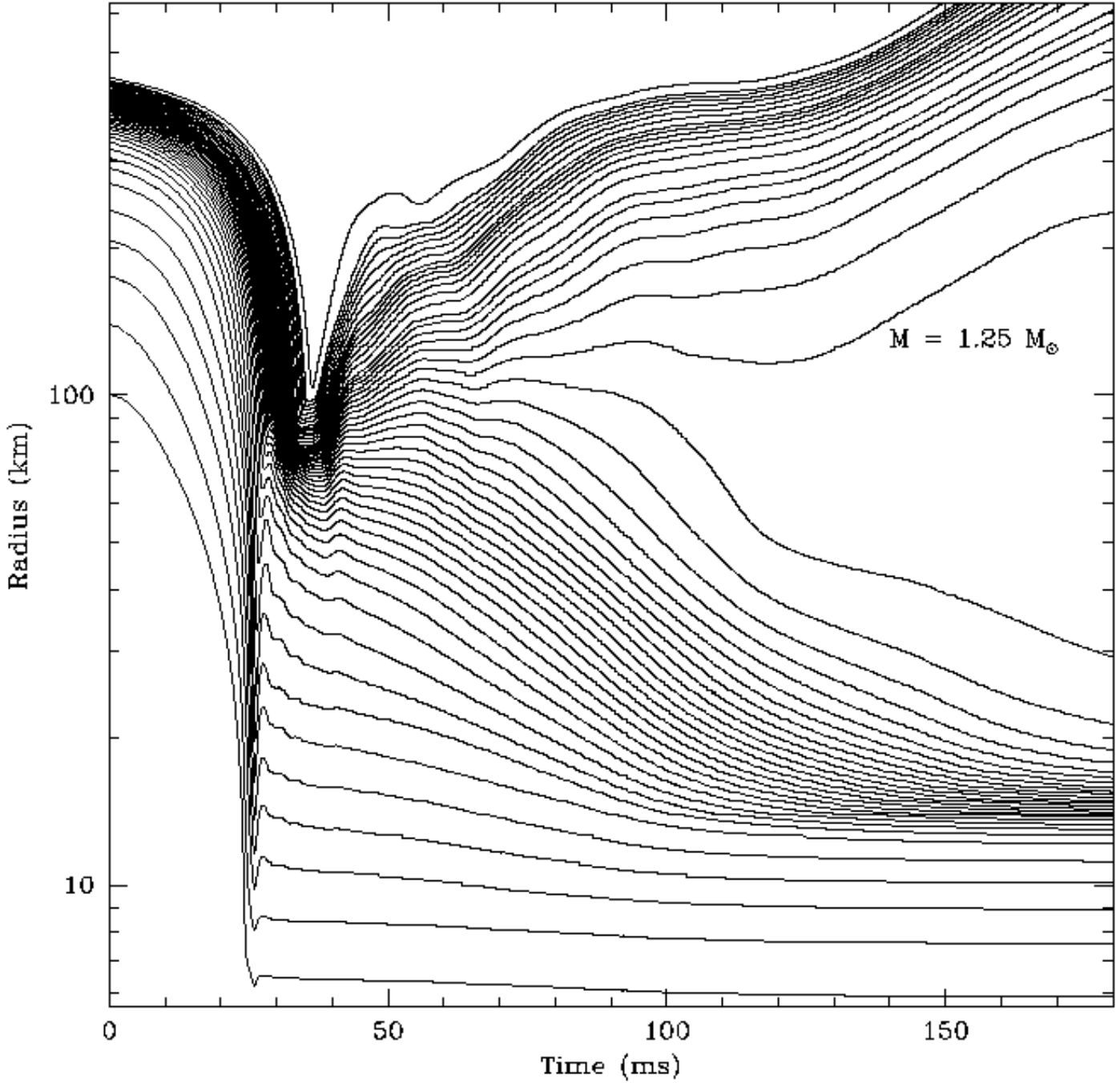}{7in}{0}{80}{80}{-250}{-50}
\caption{Mass-point trajectories for a simulation using the SL 
EOS.  The resolution is increased near the transition between ejected 
material and proto-neutron star material.  The inner 9 lines represent 
the $1 M_{\odot}$ core.}
\end{figure}

\begin{figure}
\plotfiddle{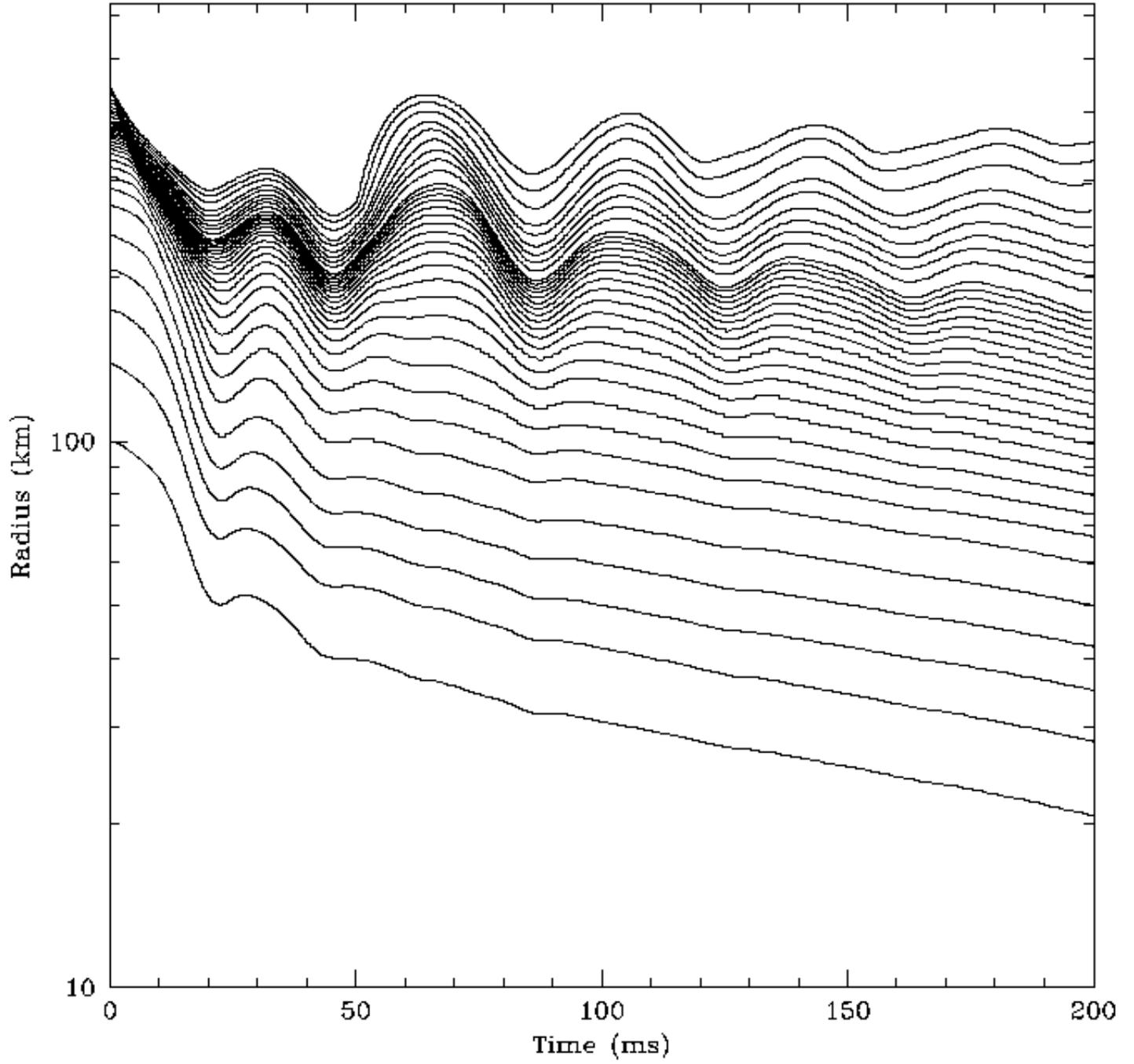}{7in}{0}{80}{80}{-250}{-50}
\caption{Identical to Figure 2 using the BCK equation of state.}
\end{figure}

\begin{figure}
\plotfiddle{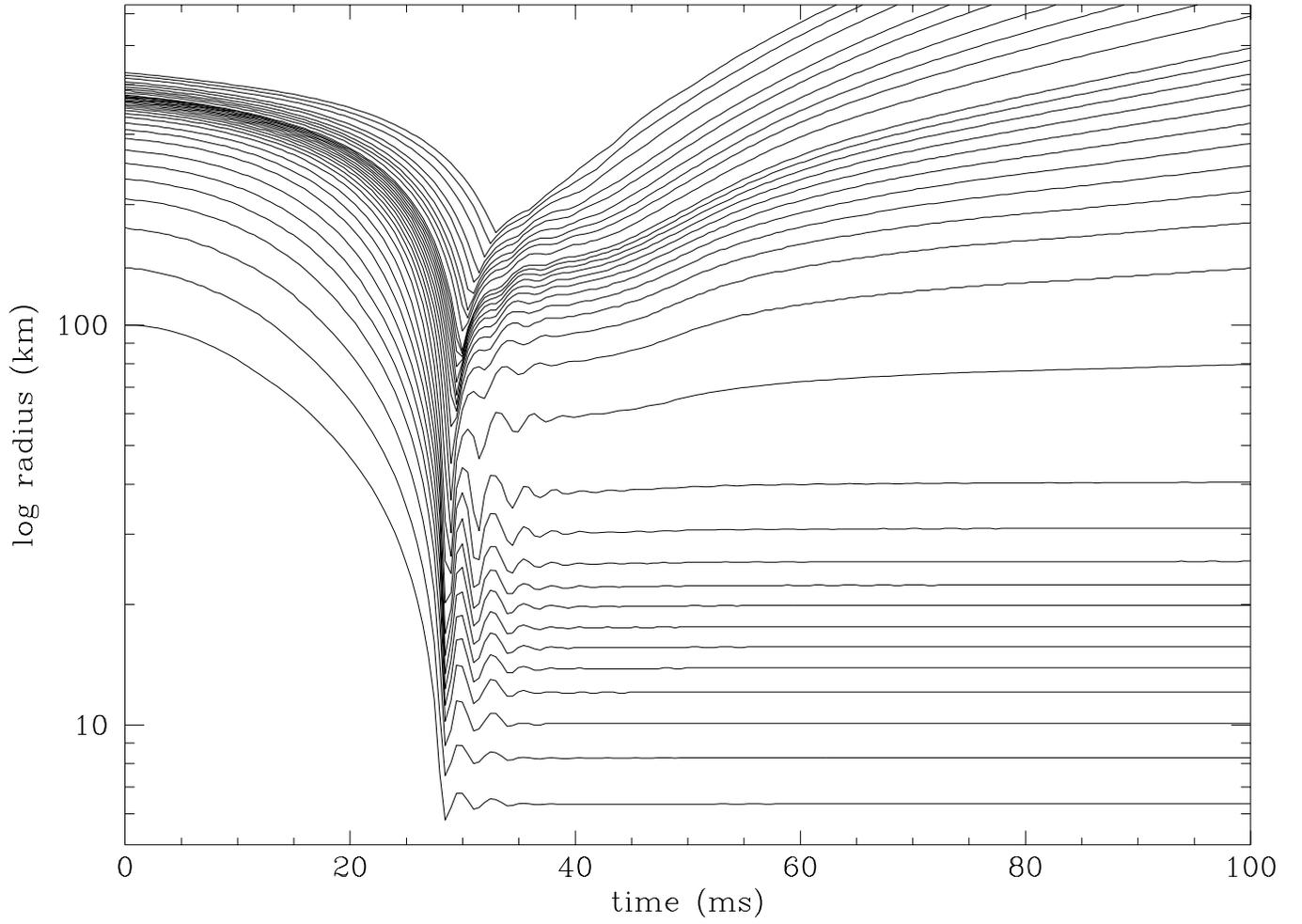}{7in}{-90}{70}{70}{-250}{500}
\caption{Identical to Figure 2 with the same SL EOS but without 
neutrino transport.  Comparison with Figure 2 differentiates a 
prompt explosion from a delayed neutrino explosion.}
\end{figure}

\begin{figure}
\plotfiddle{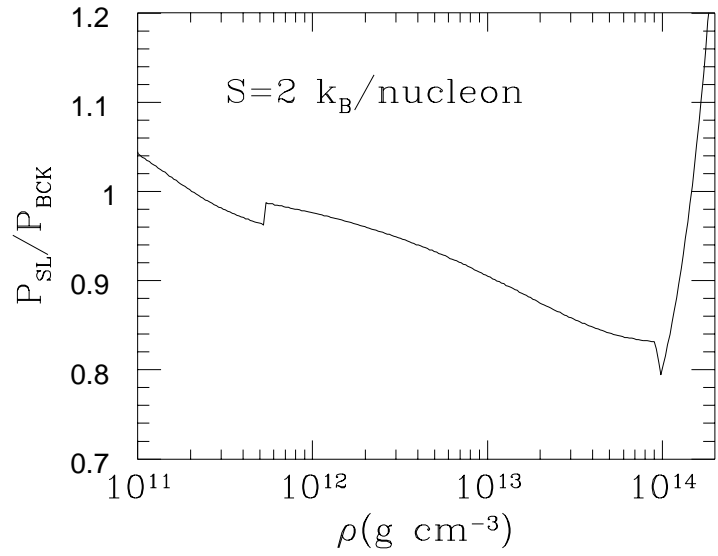}{7in}{-90}{80}{80}{-160}{320}
\caption{Ratio of pressure from the SL EOS and the BCK EOS.  
For densities less than $10^{14}$ g cm$^{-3}$, the BCK pressure is 
higher by 10-20\%.  Ye, KS}
\end{figure}

\begin{figure}
\plotfiddle{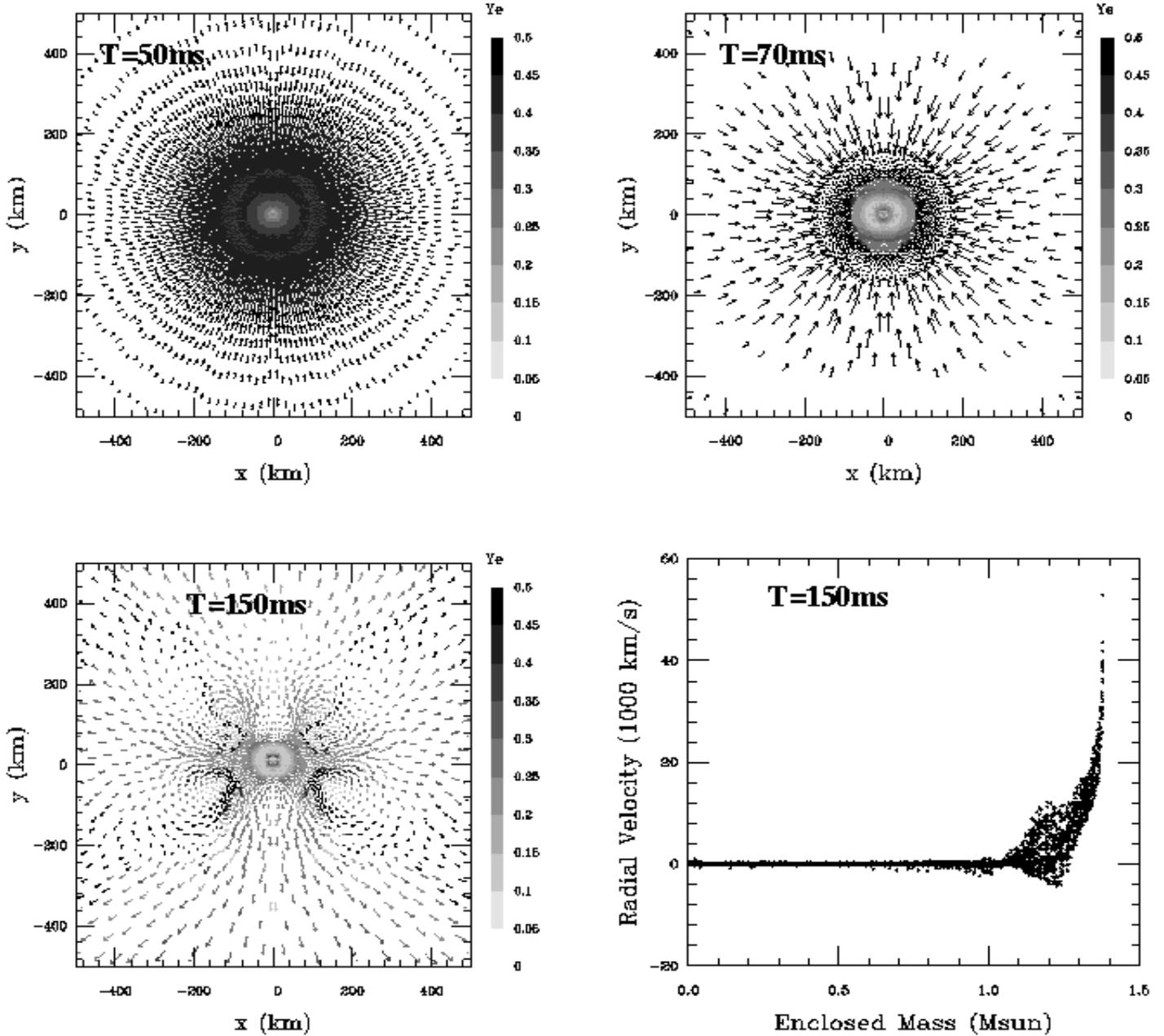}{7in}{0}{80}{72}{-250}{-50}
\caption{Time evolution of the collapse of a rotating (20\,ms) 
white dwarf.  At 50\,ms, the core of the white dwarf has 
already begun to rebound, but by 70\,ms, the shock has stalled 
and an accretion shock has formed.  Note that the material falling 
along the equator (x-axis) is further out than the material infalling 
along the rotation axis (y-axis).  At 150\,ms, the explosion 
has been launched.  Although there is some convection, 
the explosion happens so quickly that it does not effect the 
ejecta significantly.  In the lower right-hand corner, 
the radial velocity is plotted versus mass at 150\,ms.  
The outer $\sim 0.2 M_\odot$ is eventually ejected.  
The actual simulation is a 180$^{\circ}$ wedge assuming 
cylindrical symmetry which have reflected about the 
vertical axis.}
\end{figure}

\end{document}